\documentclass[12pt,twoside]{article} 

\usepackage[T1]{fontenc}
\usepackage{amsmath,amssymb,amsthm}
\usepackage{mathrsfs}
\usepackage{savesym}
\savesymbol{amalg}
\usepackage{dsfont}
\usepackage{bm}
\usepackage{multirow}
\usepackage[textwidth=2.0cm,textsize=scriptsize]{todonotes}
\usetikzlibrary{arrows}
\usepackage{subfigure}
\usepackage{tikz}
\usepackage{float}
\usetikzlibrary{shapes.geometric}
\usepackage{etoolbox}

\newcommand{\U}{\mathbf{U}}
\newcommand{\Snc}{\mathbf{S}}

\newcommand{\R}{\mathbf{R}}

\newcommand{\Nat}{\mathbb{N}}

\newcommand{\X}[1]{\mathbf{X}\left(#1\right)}
\newcommand{\Y}[1]{\mathbf{Y}\left(#1\right)}

\newcommand{\iFF}{\Leftrightarrow}

\newcommand{\zot}{$\mathds{Z}$ot}
\newcommand{\G}[1]{\mathbf{G}\left(#1\right)}

\newcommand{\F}[1]{\mathbf{F}\left(#1\right)}

\newcommand{\pair}[2]{(#1, #2)}

\makeatletter
\def\Eqlfill@{\arrowfill@\Relbar\Relbar\Relbar}
\newcommand{\longmodels}[1][]{\,|\!\!\!\ext@arrow 0359\Eqlfill@{#1}}
\makeatother

\newcommand{\set}[1]{\{ #1 \}}

\def\NOTRaisingEdge{{%
    \setbox0\hbox{\RaisingEdge}%
    \rlap{\hbox to \wd0{\hss/\hss}}\box0
}}

\def\NOTFallingEdge{{%
    \setbox0\hbox{\FallingEdge}%
    \rlap{\hbox to \wd0{\hss/\hss}}\box0
}}

\def\NOTShortPulseHigh{{%
    \setbox0\hbox{\ShortPulseHigh}%
    \rlap{\hbox to \wd0{\hss/\hss}}\box0
}}

\def\NOTShortPulseLow{{%
    \setbox0\hbox{\ShortPulseLow}%
    \rlap{\hbox to \wd0{\hss/\hss}}\box0
}}

  \DeclareFontFamily{U}  {MnSymbolA}{}
  \DeclareSymbolFont{MnSyA}         {U}  {MnSymbolA}{m}{n}
  \SetSymbolFont{MnSyA}       {bold}{U}  {MnSymbolA}{b}{n}
\DeclareFontShape{U}{MnSymbolA}{m}{n}{
    <-6>  MnSymbolA5
   <6-7>  MnSymbolA6
   <7-8>  MnSymbolA7
   <8-9>  MnSymbolA8
   <9-10> MnSymbolA9
  <10-12> MnSymbolA10
  <12->   MnSymbolA12}{}
\DeclareFontShape{U}{MnSymbolA}{b}{n}{
    <-6>  MnSymbolA-Bold5
   <6-7>  MnSymbolA-Bold6
   <7-8>  MnSymbolA-Bold7
   <8-9>  MnSymbolA-Bold8
   <9-10> MnSymbolA-Bold9
  <10-12> MnSymbolA-Bold10
  <12->   MnSymbolA-Bold12}{}
  \DeclareMathSymbol{\leftfilledspoon}{\mathrel}{MnSyA}{114}
  \DeclareMathSymbol{\leftspoon}{\mathrel}{MnSyA}{106}
  \DeclareMathSymbol{\upfilledspoon}{\mathrel}{MnSyA}{113}
  \DeclareMathSymbol{\upspoon}{\mathrel}{MnSyA}{105}
  \DeclareMathSymbol{\leftfree}{\mathrel}{MnSyA}{130}

  \DeclareFontFamily{U}  {MnSymbolB}{}
  \DeclareSymbolFont{MnSyB}         {U}  {MnSymbolB}{m}{n}
  \SetSymbolFont{MnSyB}       {bold}{U}  {MnSymbolB}{b}{n}
\DeclareFontShape{U}{MnSymbolB}{m}{n}{
    <-6>  MnSymbolB5
   <6-7>  MnSymbolB6
   <7-8>  MnSymbolB7
   <8-9>  MnSymbolB8
   <9-10> MnSymbolB9
  <10-12> MnSymbolB10
  <12->   MnSymbolB12}{}
\DeclareFontShape{U}{MnSymbolB}{b}{n}{
    <-6>  MnSymbolB-Bold5
   <6-7>  MnSymbolB-Bold6
   <7-8>  MnSymbolB-Bold7
   <8-9>  MnSymbolB-Bold8
   <9-10> MnSymbolB-Bold9
  <10-12> MnSymbolB-Bold10
  <12->   MnSymbolB-Bold12}{}
  \DeclareMathSymbol{\nleftfilledspoon}{\mathrel}{MnSyB}{114}

\DeclareRobustCommand{\shortf}[3] 
{\ensuremath{%
	 \ifthenelse{\not \equal{#3}{} }  {\stackrel{#2}{#1}_{#3}}  {\stackrel{#2}{#1}}
  }%
}%

\DeclareRobustCommand{\LogOp}[3] 
{\ensuremath{%
	 \ifthenelse{\not \equal{#2}{}}  {\mathbf{#1}_{#2}}  {\mathbf{#1}}
	 \ifthenelse{\not \equal{#3}{}}  {\!\left({#3}\right)} {}
  }%
}%

\DeclareRobustCommand{\LogOpPast}[3] 
{\ensuremath{%
	 \ifthenelse{\not \equal{#2}{}}  {\overleftarrow{\mathbf{#1}}_{#2}}  {\mathbf{#1}}
	 \ifthenelse{\not \equal{#3}{}}  {\!\left({#3}\right)} {}
  }%
}%

\DeclareRobustCommand{\LogOpInfix}[4] 
{\ensuremath{%
	 \ifthenelse{\not \equal{#3}{}}  {#3}  {}
	 \ifthenelse{\not \equal{#2}{}}  {\mathbf{#1}_{#2}}  {\mathbf{#1}}
	 \ifthenelse{\not \equal{#4}{}}  {#4} {}
  }%
}%

\date{} 

\begin{document} 






\centerline {\Large{\bf LTL-based Verification of Reconfigurable Workflows}} 

\centerline{} 

\centerline{\bf {Manuel Mazzara}} 

\centerline{Innopolis University, Russia} 


\centerline{} 

\centerline{}

\begin{abstract} Logics and model-checking have been successfully used in the last decades for modeling and verification of various types of hardware (and software) systems. While most languages and techniques emerged in a context of monolithic systems with a limited self-adaptability, modern systems require approaches able to cope with dynamically changing requirements and emergent behaviors. The emphasis on system reconfigurability has not been followed by an adequate research effort, and the current state of the art lacks logics and model checking paradigms that can describe and analyze complex modern systems in a comprehensive way. This paper describes a case study involving the dynamic reconfiguration of an office workflow. We state the requirements on a system implementing the workflow and its reconfiguration and we prove workflow reconfiguration termination by providing a compilation of generic workflows into LTL, using the Bound model checker \zot{}. The objective of this paper is demonstrating how temporal logics and model checking are effective in proving properties of dynamic, reconfigurable and adaptable systems. This simple case study is just a "proof of concept" to demonstrate the feasibility of our ideas.

\end{abstract} 



\section{Introduction}
\label{section-intro}

Dependable systems have been investigated for decades \cite{Avizienis:2004}, being software reliability of primary importance not only (and not in particular) for systems of everyday use, like operating systems or software for text and video editing or music players, but especially for life-critical systems, like those in automotive and aviation systems. However, modern systems impose new challenges to system and software engineers as well as to the overall computer science community. These days we increasingly build systems that respond to external stimuli and change their computational behavior accordingly. In such situations, their reconfigurations are externally determined. This poses reconfiguration and adaptability requirements and the need to evaluate system dependability from a broader perspective than before.

The next generation of dependable systems is expected to have significant evolution requirements \cite{kn:CoyEtAl10} and it is highly likely that the system will have to be redesigned (i.e. reconfigured)  during its lifetime (possibly more than once) in order to meet new requirements. Therefore, it will be impossible to foresee all the requirements that a system will have to meet in future when it is being designed \cite{kn:MenMagRum10}. In particular, certain classes of dependable systems, such as control systems, must be dynamically reconfigured \cite{kn:KarMasOstSch10}, because it is unsafe or impractical or too expensive to do otherwise. The dynamic reconfiguration of a system is defined as the change at runtime of the structure of the system -- consisting of its components and their communication links -- or the hardware location of its software components \cite{kn:Bha13}.

This paper focuses on dynamic software reconfiguration, because software is much more mutable than hardware \cite{Mazzara2010, AbouzaidMMD12}. However, several techniques that have been applied to hardware verification can still be applied to software. Logics and model-checking, for example, have been successfully used in the last decades for modeling and verification of various types of hardware (and software) systems. While most languages and techniques emerged in a context of monolithic systems with a limited self-adaptability, modern systems require approaches able to cope with dynamically changing requirements and emergent behaviors. The emphasis on system reconfigurability has not been followed by an adequate research effort, and the current state of the art lacks logics and model checking paradigms that can describe and analyze complex modern systems in a comprehensive way. 

In this work, we tackle the open problem of verifying reconfigurable systems (and in particular reconfigurable workflows starting from \cite{Mazzara06}) by temporal logics and model checking. This paper describes a case study involving the dynamic reconfiguration of an office workflow similar to the one studied in \cite{MazzaraADB11}. The role of temporal logics in verification and validation is two-fold. First, temporal logic allows abstract, concise and convenient expression of required properties of a system. Linear Temporal Logic (LTL) is often used with this goal in the verification of finite-state models, e.g., in model checking~\cite{BK08}. Second, temporal logic allows a descriptive approach to specification and modeling (see, e.g.,~\cite{MS94,FMMR12}). A descriptive model is based on axioms, written in some (temporal) logic, defining  a system by means of its general properties, rather than by an operational model based on some kind of machine behaving in the desired way. In this case, verification typically consists of satisfiability checking of the conjunction of the model and of the (negation of) its desired properties. An example of the latter approach is Bounded Satisfiability Checking (BSC) \cite{PMS12}, where Metric Temporal Logic (MTL) specifications on {\em discrete} time and properties are translated into Boolean logic, in an approach similar to Bounded Model Checking of LTL properties of finite-state machines.

The rest of the paper is organized as follows: Section \ref{section-casestudy} describes the case study, in which a simple office workflow for order processing is reconfigured. The reconfiguration 
of the workflow is modelled and analyzed using temporal logic and, in particular, in Section      
\ref{section-ltl} a general encoding of workflows into LTL is provided. In Section \ref{section-experimental} the implementation of this translation is illustrated and tests have been carried out to validate its correctness. Finally, Section \ref{section-conclusions} draw conclusive remarks and focus on future developments of this work.

\section{Case study}
\label{section-casestudy}

The case study approached in this paper describes dynamic reconfiguration of 
an office workflow for order processing that is commonly found in large and 
medium-sized organizations \cite{Ellis:1995}. These workflows typically handle 
large numbers of orders. Furthermore, the organizational environment of a workflow, 
can change in structure, procedures, policies and legal obligations in a manner 
unforeseen by the original designers of the workflow. Therefore, it is necessary 
to support the unplanned change of these workflows. Furthermore, the state of an
order in the old configuration may not correspond to any state of the order in the 
new configuration. These factors, taken in combination, imply that instantaneous 
reconfiguration of a workflow is not always possible; neither is it practical to 
delay or abort large numbers of orders because the workflow is being reconfigured. 
The only other possibility is to allow overlapping modes for the workflow during 
its reconfiguration.

\begin{figure}[H]
\centering
\includegraphics[scale=0.5]{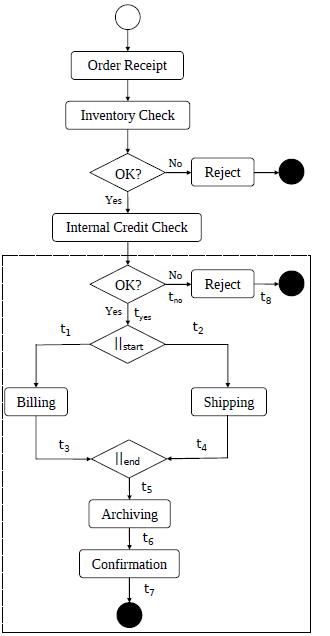}
\caption{First configuration}
\label{fig:first}
\end{figure}

The workflow consists of a graph of activities and the configuration is the structure 
of this graph. Initially, the workflow executes in Configuration 1 and must satisfy 
the requirements on Configuration 1. Subsequently, the workflow must be reconfigured 
through a process to Configuration 2 in a way that the requirements on the reconfiguration 
process and on Configuration 2 are satisfied. The full details of the two configurations 
and the relative requirements are given in \cite{MazzaraADB11}. The initial configuration 
of the workflow is depicted in Figure \ref{fig:first}.

\begin{figure}[H]
\centering
\includegraphics[scale=0.5]{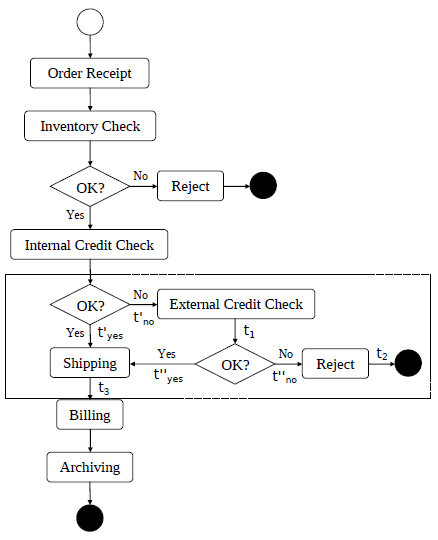}
\caption{Second configuration}
\label{fig:second}
\end{figure}

After some time, the management of the organization using the workflow decides to change it
in order to increase opportunities for sales, improve the synchronisation between \texttt{Billing} and \texttt{Shipping}, and to simplify the workflow. The new configuration of the workflow is Configuration 2 as depicted in Figure \ref{fig:second}. The two pictures below represent a reconfiguration between the two workflows. Dashed rectangles are the portions of the workflows that are translated in Section~\ref{section-ltl}.

In order to achieve a smooth transition from Configuration 1 to Configuration 2 of the workflow,
the process of reconfiguration must meet the following requirements:

\begin{enumerate}
   \item Reconfiguration of a workflow should not necessarily result in the rejection of an order
   \item Any order being processed that was accepted \textbf{before} the start of the reconfiguration must satisfy
         all the requirements on Configuration 1
   \item Any order accepted \textbf{after} the start of the reconfiguration must satisfy all the requirements on Configuration 2
   \item The reconfiguration process must terminate
\end{enumerate}

\subsection*{Methodology}

\begin{figure}[H]
\centering
\includegraphics[scale=0.7]{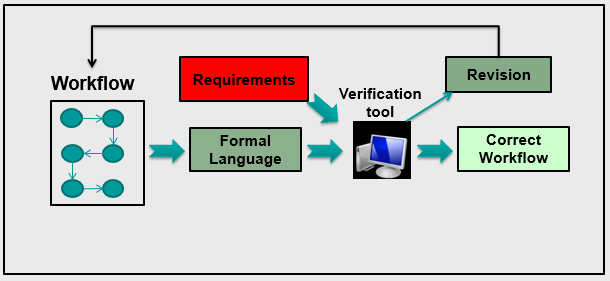}
\caption{Workflow Verification Process}
\label{fig:verification}
\end{figure}

The process of verification is depicted in Figure \ref{fig:verification}. In this work, we provide an instrument to designers for workflows revision, i.e. a procedure to follow until the requirements are finally met. To do this, we encode the system by means of a formal language, as discussed in Section \ref{section-ltl}, and we automatically determine its correctness as shown in Section \ref{section-experimental}. Differently from several others formal approaches that only offer languages without a method to apply them (for a detailed discussion see \cite{Mazzara2010bis}), this methodology, together with software tools, aims at offering a complete practical toolkit for software and systems engineers working in the field of workflow reconfiguration. 

\section{Verification by LTL specification}\label{section-ltl}

LTL~\cite{LPZ85} is one of the most popular descriptive languages for defining temporal behaviors that are represented as sequences of observations.  The time model adopted in this logic is a totally ordered set (e.g., $(\mathbb{N},<)$) whose elements are the positions where the behavior is observed. 
LTL allows the expression of positional orders of events both towards the past and the future.
Let $AP$ be a finite set of atomic propositions.
Well-formed LTL formulae are defined as follows:
\begin{equation*}
  \phi :=
  \begin{gathered}
    p \mid \phi \wedge \phi \mid \neg \phi \mid   \X{\phi} \mid \Y{\phi} \mid \phi\U\phi \mid \phi\Snc\phi
  \end{gathered}
\end{equation*}
where $p \in AP$, $\mathbf{X}$, $\mathbf{Y}$, $\U$ and $\Snc$ are the usual ``next'', ``previous'', ``until'' and ``since'' modalities.
The dual operator ``release'' $\R$  is defined as usual, i.e., $\phi\R \psi$ is $\neg(\neg \phi \U \neg\psi)$ and the ``Globally" $\mathbf{G}\phi$ operator is  $\mathit{false} \R \phi$.

The semantics of LTL formulae is defined with respect to a strict linear order representing time $\pair{\Nat}{<}$.
Truth values of propositions in $AP$ are defined by function $\pi :\Nat \to \wp(AP)$ associating a subset of the set of propositions with each element of $\Nat$.
The semantics of a LTL formula $\phi$ at instant $i\geq 0$
over a linear structure $\pi$ is recursively defined as in Table~\ref{tab:LTLsemantics},
\begin{table}[h!]
\begin{equation*}
\begin{aligned}
\pi, i \models p  &\iFF  p \in \pi(i) \text{ for } p \in AP \\
\pi, i \models \X \phi &\iFF \pi,i+1 \models \phi \\
\pi, i \models \Y \phi &\iFF \pi,i-1 \models
\phi \wedge i>0\\
\pi, i \models \phi\U\psi &\iFF
\exists \, j\geq i: \pi,j \models \psi \ \wedge \pi,n \models \phi \ \forall \ i\leq n < j \\
\pi, i \models \phi\Snc\psi &\iFF
\exists \, 0\leq j \leq i: \pi,j \models \psi \, \wedge \pi,n \models \phi \ \forall \ j < n \leq i \\ 
\end{aligned}
\end{equation*}
\caption{Semantics of CLTL. Boolean connectives are omitted for brevity.}\label{tab:LTLsemantics}
\end{table}
A formula $\phi \in$ CLTL is \emph{satisfiable} if there exists 
a pair $\pi$ such that $\pi,0 \models \phi$. 

\begin{table}
\small
\centering
\label{tab:translation}
\begin{tabular}{c|c|c}
$[A]_{t_{out}(A)}$ &
\begin{minipage}{7cm}
\begin{gather}
A \Rightarrow (A \land \neg t_{\mathrm{out}}(A)) \U (t_{\mathrm{out}}(A))\label{f-act-followed-trans-1} \\
\bigwedge_{i\in \mathit{out}(A)} (t_i \Rightarrow \Y{A} \land \neg A)\label{f-act-followed-trans-2} %
\end{gather}
\end{minipage} & 
\begin{minipage}{2cm}
\begin{tikzpicture}
\draw (0,0) node[rectangle,draw] {$A$};
\draw[->]  (0.47,0) -- (1,0);
\draw[->]  (-0.47,0) -- (-1,0);
\draw[->]  (0,-0.5) -- (0,-1);
\draw  (0.8,0.2) node {$t_2$};
\draw (-0.8,0.2) node {$t_1$};
\draw (0.2,-0.8) node {$t_i$};
\end{tikzpicture}
\end{minipage}\\
\hline 
${_{t_{in}(A)}}[A]$ & 
\begin{minipage}{7cm}
\begin{gather}
A \Rightarrow (A \land \neg t_{\mathrm{in}}(A)) \Snc (t_{\mathrm{in}}(A))\label{f-trans-followed-act-1} \\
\bigwedge_{i\in \mathit{in}(A)} (t_i \Rightarrow \X{A} \land \neg A)\label{f-trans-followed-act-2}
\end{gather}
\end{minipage} & 
\begin{minipage}{2cm}
\begin{tikzpicture}
\draw (0,0) node[rectangle,draw] {$A$};
\draw[<-]  (0.47,0) -- (1,0);
\draw[<-]  (-0.47,0) -- (-1,0);
\draw[<-]  (0,0.5) -- (0,1);
\draw  (0.8,0.2) node {$t_2$};
\draw (-0.8,0.2) node {$t_1$};
\draw (0.2,0.8) node {$t_i$};
\end{tikzpicture}
\end{minipage}\\
\hline
$[\cdot+\cdot]$ & 
\begin{minipage}{7cm}
\begin{gather}
t_1 \Rightarrow \neg t_2 \label{f-cond-1}\\
\oplus \Rightarrow \neg \Y{\oplus} \land \neg \X{\oplus}\label{f-cond-2}
\end{gather}
\end{minipage} & 
\begin{minipage}{2cm}
\begin{tikzpicture}
\draw (0,0) node[diamond,draw] {+};
\draw[->]  (0.4,0) -- (1,0);
\draw[->] (-0.4,0) -- (-1,0);
\draw (0.8,0.2) node {$t_1$};
\draw (-0.8,0.2) node {$t_2$};
\end{tikzpicture}
\end{minipage}\\
\hline
$[\cdot\mid \cdot]$ & 
\begin{minipage}{7cm}
\begin{gather}
\bigwedge_{t_1,t_2 \in \mathit{out}(\rVert)} (t_1 \iFF  t_2) \text{ or }\bigwedge_{t_1,t_2 \in \mathit{in}(\rVert)} (t_1 \iFF  t_2)\label{f-split-1}\\
\rVert \Rightarrow \neg \Y{\rVert} \land \neg \X{\rVert}\label{f-split-2}
\end{gather}
\end{minipage} & 
\begin{minipage}{5cm}
\vspace{10pt}
\begin{tabular}{cc}
\begin{tikzpicture}
\draw (0,0) node[diamond,draw] {$\rVert$};
\draw[->]  (0.47,0) -- (1,0);
\draw[->]  (-0.47,0) -- (-1,0);
\draw[->]  (0,-0.5) -- (0,-1);
\draw  (0.8,0.2) node {$t_2$};
\draw (-0.8,0.2) node {$t_1$};
\draw (0.2,-0.8) node {$t_i$};
\end{tikzpicture}
&
\begin{tikzpicture}
\draw (0,0) node[diamond,draw] {$\rVert$};
\draw[<-]  (0.47,0) -- (1,0);
\draw[<-]  (-0.47,0) -- (-1,0);
\draw[<-]  (0,0.5) -- (0,1);
\draw  (0.8,0.2) node {$t_2$};
\draw (-0.8,0.2) node {$t_1$};
\draw (0.2,0.8) node {$t_i$};
\end{tikzpicture}
\end{tabular}
\end{minipage}
 \\
\end{tabular} 
\caption{Workflow LTL encoding}
\end{table}

Workflows model execution of systems as sequences of activities.
Transitions, conditional choices and parallel ``split" interleave the activities and determine uniquely the flow of the execution, i.e., the sequence of activities that realizes the computation.
An activity is an abstraction of a compound of actions that are performed by the real systems. 
Although they can be modelled as atomic computations, we adopt a different perspective for which the activities, being actions in the real world, have a non-punctual duration.
To translate workflows into an LTL formula, we assume that an activity is always followed by a transition, and viceversa,
and that conditional tests and splits are special activities that have punctual duration.
When an activity is performed, the firing of the outgoing transition lets the system change allowing it to execute the next activity.
Therefore, our LTL translation allows modelling of workflows as sequences of activities and transitions, in strict alternation.

With no loss of generality, we assume that no element in the same graph is duplicated.
By this assumption, we associate each element with an atomic proposition that uniquely identifies it.
Observe, however, that there could be some activity that are shared by the two configurations but each transition has a unique name and no transition with the same label appears twice.
We write $A$ to represent the activity $A$ in the workflow, while we write $t$ to indicate a transition between two activities.
If $A$ holds at position $i$ then the workflow is performing activity $A$ at that position; similarly for $t$.
We introduce $\rVert$ and $\oplus$ to indicate the split and the conditional activity, respectively; $\mathit{start}$ and $\mathit{end}$ to indicate two special activities representing the starting and the final activity of the workflow.
Workflow diagrams are translated according to rules in Table 2.

Let $A$ be an activity, $\mathit{out}(A)$ be the set of outgoing transition starting from $A$ and $\mathit{in}(A)$ be the set of ingoing transition leading to $A$.
By definition, a correct workflow is such that $\mathit{out}(A)\geq 1$, for all activity $A$, except for activity $\mathit{end}$ and that $\mathit{out}(A)\geq 1$, for all activity $A$, except for activity $\mathit{start}$.
Let $t_{\mathrm{out}}(A)$ be the disjunction $\bigvee_{t_i \in \mathit{out}(A)} t_i$ and $t_{\mathrm{in}}(A)$ be the disjunction $\bigvee_{t_i \in \mathit{in}(A)} t_i$.

Formula~\eqref{f-act-followed-trans-1} states that if activity $A$ holds at the current position, then it is true, at that position, that $A$ but not $t_{\mathrm{out}}(A)$ holds until at least one transition in $\mathit{out}(A)$ holds.
In this way, activity $A$ lasts until one of its outgoing transition fires.
Formula~\eqref{f-act-followed-trans-2} imposes that if transition $t_i \in \mathit{out}(A)$ holds at position $i$, then in the previous one $i-1$, activity $A$ holds.
This necessary condition enforces that a transition fires only if the activity from which it originates has just been terminated.

Formula~\eqref{f-trans-followed-act-1} states that if activity $A$ holds at the current position, then it is true, at that position, that $A$ but not $t_{\mathrm{in}}(A)$ holds since at least one transition in $\mathit{in}(A)$ has been fired.
In this way, activity $A$ has lasted since one of its ingoing transition fired.
Formula~\eqref{f-trans-followed-act-2} imposes that if transition $t_i \in \mathit{in}(A)$ holds at position $i$, then, in the next one $i+1$, activity $A$ holds.
This necessary condition enforces that a transition fires only if the activity to which it leads will be performed in the next position.

The translation for $[A]_{t_{out}(A)}$ and ${_{t_{in}(A)}}[A]$ does not impose constraints on transitions in $t_{out}(A)$ and $t_{in}(A)$.
In the case of $[A]_{t_{out}(A)}$, the execution flow after an activity $A$ is nondeterministically defined as it is determined by at least one transition that fires when $A$ terminates.
Simmetrically, the execution of an activity $A$ that is reached by more than one transition in ${_{t_{in}(A)}}[A]$ can be started by at least one execution that is active before $A$. 
The conditional block $[\cdot + \cdot]$ and the split $[\cdot\mid\cdot]$ are modelled by using the rules (1)-(4) and, in addition, specific constraints to enforce the proper flow of execution.
To model the flow of the conditional block, we force the execution of the two branches to be exclusive as for the \textit{if-then-else} construct.
The translation for the split block is similar yet it enforces the synchronization of all the transitions starting from, or yielding to, the activity split.

The conditional block $[\cdot\mid\cdot]$ is translated compositionally.
For each conditional block we introduce a new fresh atomic proposition $\oplus$.
Formulae \eqref{f-act-followed-trans-1}, \eqref{f-act-followed-trans-2}, \eqref{f-trans-followed-act-1} and \eqref{f-trans-followed-act-2} rule the semantics of the block as sequence of actions.
In addition, Formula \eqref{f-cond-1} impose that only one branch is executed, by forcing the strict complementarity between $t_1$ and $t_2$.
Formula \eqref{f-cond-2} enforces punctuality of $\oplus$ and states that if $\oplus$ holds at position $i$ then in the next and in the previous positions it does not hold.

The translation for the split block is similar to the one defined for the conditional block.
For each split block we introduce a new fresh atomic proposition $\rVert$.
Formulae \eqref{f-act-followed-trans-1}, \eqref{f-act-followed-trans-2}, \eqref{f-trans-followed-act-1} and \eqref{f-trans-followed-act-2} rule the semantics of the block as sequence of actions and Formula \eqref{f-cond-2} enforces punctuality, similarly to the previous case.
The only difference is Formula \eqref{f-split-1} that is divided into two parts that are used exclusively.
Both of them impose strict contemporaneity of all the transitions involved in activity $\rVert$.
The first one is defined only for the initial point of a split block and states that all the outgoing transitions starting from it, occur at the same time.
The second one is similar but for the final point of a split block, where all the parallel computations must join before proceeding further.
It states that all the ingoing transitions leading to it, occur at the same time.

\newcommand{\ship}{\mathit{Ship}}
\newcommand{\bill}{\mathit{Bill}}
\newcommand{\arch}{\mathit{Arch}}
\newcommand{\conf}{\mathit{Conf}}
\newcommand{\rej}{\mathit{Rej}}
\newcommand{\eendwf}{\mathit{end}} 

For the sake of brevity, we provide, as an example, the translation of a portion of the 
two workflows of Figure~\ref{fig:first} and \ref{fig:second}.
Let us consider the first one \ref{fig:first} from the second conditional.
We introduce the following atomic propositions representing the activities.
We use $\oplus$ for modelling the conditional block, $\rej$ for \textit{Reject}, $\rVert_{start}$ and $\rVert_{end}$ for the activities defining the starting and the ending point of the join block, $\arch$, $\ship$, $\bill$ and $\conf$ for \textit{Archiving}, \textit{Shipping}, \textit{Billing} and \textit{Confirmation}.
Finally, we introduce the final state $\eendwf$.
The following formulae model the portion of the workflow.
We unify the translation for activities $\ship$, $\ship$, $\bill$ and $\conf$, being similar.
Let $A\in \set{\ship,\arch,\bill,\conf,\rej}$ and $t'\in\set{t_1,t_2,t_5,t_6,t_{no}}$, $t''\in\set{t_3,t_4,t_6,t_7,t_8}$.
We use the following cases for $(A,t',t'')$: $(\ship,t_2,t_4)$, $(\bill,t_1,t_3)$, $(\arch,t_5,t_6)$, $(\conf,t_6,t_7)$ and $(\rej,t_{no},t_8)$.

\scriptsize
\[
\begin{array}{ccc}
\begin{gathered}
\oplus \Rightarrow (\oplus \land \neg (t_{yes}\lor t_{no})) \U (t_{yes}\lor t_{no})) \\
t_{yes} \Rightarrow \Y{\oplus} \land \neg \oplus\\
t_{no} \Rightarrow \Y{\oplus} \land \neg \oplus\\
t_{yes} \Rightarrow \neg t_{no}\\ 
\oplus \Rightarrow \neg\Y{\oplus} \land \neg\X{\oplus}
\end{gathered} & 
\begin{gathered}
\rVert_{start} \Rightarrow (\rVert_{start} \land \neg t_{yes}) \Snc \; t_{yes}\\
t_{yes} \Rightarrow \X{\rVert_{start}} \land \neg \rVert_{start}\\
\rVert_{start} \Rightarrow (\rVert_{start} \land \neg (t_1\lor t_2)) \U (t_2\lor t_2)) \\
t_2 \Rightarrow \Y{\rVert_{start}} \land \neg \rVert_{start}\\
t_1 \Rightarrow \Y{\rVert_{start}} \land \neg \rVert_{start}\\
t_1 \iFF t_2\\
\rVert_{start} \Rightarrow \neg\Y{\rVert_{start}} \land \neg\X{\rVert_{start}}
\end{gathered} &
\begin{gathered}
A \Rightarrow (A \land \neg t') \Snc \; t'\\
t' \Rightarrow \X{A} \land \neg A\\
A \Rightarrow (A \land \neg t'') \U \; t'') \\
t'' \Rightarrow \Y{A} \land \neg A
\end{gathered}
\\  \\
\begin{gathered}
\rVert_{end} \Rightarrow (\rVert_{end} \land \neg (t_3\lor t_4)) \Snc \; (t_3\lor t_4)\\
t_4 \Rightarrow \X{\rVert_{end}} \land \neg \rVert_{end}\\
t_3 \Rightarrow \X{\rVert_{end}} \land \neg \rVert_{end}\\
\rVert_{end} \Rightarrow (\rVert_{end} \land \neg t_5) \U \; t_5) \\
t_5 \Rightarrow \Y{\rVert_{end}} \land \neg \rVert_{end}\\
t_3 \iFF t_4\\
\rVert_{end} \Rightarrow \neg\Y{\rVert_{end}} \land \neg\X{\rVert_{end}}
\end{gathered} &
\begin{gathered}
\eendwf \Rightarrow (\eendwf \land \neg (t_7\lor t_8)) \Snc \; (t_7\lor t_8)\\
t_7 \Rightarrow \X{\eendwf} \land \neg \eendwf\\
t_8 \Rightarrow \X{\eendwf} \land \neg \eendwf
\end{gathered} &
\begin{gathered}
\end{gathered} 
\end{array} 
\]
\newcommand{\ecc}{\mathit{Ecc}}

\normalsize

The partial translation of the second workflow follows.
In particular, only the second and the third conditional block and the activities \textit{External Credit Check} ($\ecc$) and \textit{Shipping} are considered.
We represent the first conditional with $\oplus'$ and the second one with $\oplus''$.
We abbreviate the notation for the two blocks with $\oplus^*$, where $*$ is the a quote $'$ or a double quote $''$, and for the associated translations with $t^*_{yes}$ and $t^*_{no}$.
The translation for $\oplus^i$ follows exactly the same structure as the one previously defined for the first workflow, so we do not provide it.

\scriptsize
\[
\begin{array}{ccc}
\begin{gathered}
\ecc \Rightarrow (\ecc \land \neg t'_{no}) \Snc \; t'_{no}\\
t'_{no} \Rightarrow \X{\ecc} \land \neg \ecc\\
\ecc \Rightarrow (\ecc \land \neg t_1) \U \; t_1) \\
t_1\Rightarrow \Y{\ecc} \land \neg \ecc
\end{gathered} & 
\begin{gathered}
\ship \Rightarrow (\ship \land \neg (t'_{yes}\lor t''_{yes})) \Snc \; (t'_{yes}\lor t''_{yes})\\
t'_{yes} \Rightarrow \X{\ship} \land \neg \ship\\
t''_{yes} \Rightarrow \X{\ship} \land \neg \ship\\
\ship \Rightarrow (A \land \neg t_3) \U \; t_3) \\
t_3 \Rightarrow \Y{t_3} \land \neg \ship
\end{gathered}
 &
\begin{gathered}
\rej \Rightarrow (\rej \land \neg t''_{no}) \Snc \; t''_{no}\\
t''_{no} \Rightarrow \X{\rej} \land \neg \rej \\
\rej \Rightarrow (\rej \land \neg t_2) \U \; t_2) \\
t_2 \Rightarrow \Y{\rej} \land \neg \rej
\end{gathered} 
\end{array} 
\] 
\normalsize
In this work, we assume that when a workflow instance terminates, it never resumes.
We enforce it through the formula $\eendwf \Rightarrow \X{\G{\neg\eendwf}}$.

The previous formulae only apply for the case of one workflow.
In our model, many workflows can be active at the same time and to represent different instances of the same workflow, we introduce as many activities $A^j$ and transitions $t^j$ as the number of workflows to be represented, where $j\in \Nat$ is the index representing the $j$-th instance.
All the formulae defining the workflow are replicated to define the behaviour for all the instances of the workflow.
To model the reconfiguration process, we need to represent when an instance is active or not.
A workflow is active at a certain position $i$ when at least one of its activity or transition holds at $i$.
We introduce for all the workflows an atomic proposition $W^j$, with $j\in \Nat$, that holds if, and only if, the previous condition is met.
Let $E^j_i$ be the set of propositions representing activities and transitions defining the $j$-th instance of the workflow for the $i$-th configuration.
The following formula constraints $W^j$:
\begin{equation}\label{f-active-wf}
W^j \iFF \bigvee_{e \in E^j_1 \cup E^j_2} e.
\end{equation}
Observe that $E^j_1 \cap E^j_2$ may, in general, be non empty.
This is the case of configurations of the same workflow that share some activity or transitions.
A configuration of a workflow is represented as the conjunction of all the formulae that translate it, that are defined by rules in Table~\ref{tab:translation}.
We introduce proposition $C^j_1$ and $C^j_2$ to capture the configuration in Figure~\ref{fig:first} and in Figure~\ref{fig:second}, for all the instances $j$.
Let $WS^j_i$ be the set of all formulae translating configuration $i \in \set{0,1}$ of (an instance of) workflow $j$ that are defined according to rules in Table~\ref{tab:translation}. 
Then, the formula defining the rules characterizing configuration $i$ is 
\begin{equation}\label{f-config}
C^j_i \Rightarrow \bigwedge_{\phi_h \in WS^j_i} \phi_h.
\end{equation}
Formula\ref{f-config} states that if $C^j_i$ holds at the current position, then all the formulae $\phi_h$ in $WS^j_i$ determine the current configuration and rule the execution flow.
Given an instance $j$, to guarantee that all, and only, the activities and transitions that appear in the first configuration may occur when $C^j_1$ holds (but not the activities and transitions that appear in the second configuration yet not in the first one), we add the following formula.
Let $\bar{E}^j_1 = E^j_2\setminus (E^j_1\cap E^j_2)$ be the set of activities and transitions that appear in the second configuration but not in the first one; and let $\bar{E}^j_2 = E^j_1\setminus (E^j_1\cap E^j_2)$ be the set of activities and transitions that appear in the first configuration but not in the first second.
\begin{equation}\label{f-uniqueness-I}
\bigwedge_{i\in \set{1,2}} C^j_i \Rightarrow (\bigvee_{e \in E^j_i} e) \land \neg (\bigwedge_{e \in \bar{E}^j_i} e)
\end{equation}

The reconfiguration is represented by proposition $R$.
We assume that there is only one reconfiguration in the system and that, from the position where it occurs, all the workflows behave as the second configuration defined in Figure~\ref{fig:second}.
Before the reconfiguration point, instance behaviour conforms to the first workflow of Figure~\ref{fig:first}.
To model the reconfiguration, being it unique, we first impose that when $R$ holds, at a certain position, then it holds forever in the future.
This is achieved by the formula $R\Rightarrow \G{R}$.
The choice of the workflow that rules the behaviour of an active instance is determined in the following way and it is defined by the next formula.
If an instance starts the execution before the point where the reconfiguration occurs then its behaviour conforms to the workflow of Figure~\ref{fig:first}.
In the other case, i.e., the workflow instance starts after the reconfiguration point, then its behaviour conforms to the workflow of Figure~\ref{fig:second}.
\begin{equation}\label{f-reconf}
\begin{gathered}
\left(
\begin{gathered}
(W^j\land \neg R) \U (\neg W^j) \ \ \lor \\
(W^j\land \neg R) \U ((W^j\land R) \U (\neg W^j\land R))
\end{gathered}
\right)
\Rightarrow C^j_1 \quad \land \\
(W^j\land R) \U (\neg W^j) \Rightarrow C^j_2 \quad \land \\
\neg\F{W^j \land \neg R} \Rightarrow \mathbf{G}(C^j_2).
\end{gathered}
\end{equation}
The first formula of the previous conjunction states that if, at some position, 
(i) the instance $j$ is active when the reconfiguration has not yet occurred until it terminates, 
(ii) the instance $j$ is active when the reconfiguration has not yet occurred until a position where $j$ is still active and the reconfiguration occurred until it terminates,
then its behaviour is defined by the first configuration of the workflow in Figure~\ref{fig:first}.
The second formula, defines the opposite situation where an instance $j$ starts after the reconfiguration, i.e., the instance $j$ is active when the reconfiguration has occurred until it terminates. 
In such case, the second workflow of Figure~\ref{fig:second} defines the behaviour for $j$.
The last formula imposes the behaviour defined by the second workflow over the whole time line if the instance $j$ never start before the reconfiguration.
The definition of the model is the conjunction of all the previous formulae, i.e., the translations of the workflows, Formula~\eqref{f-active-wf}, Formula~\eqref{f-config} and Formula~\eqref{f-reconf}, globally quantified over time by $\mathbf{G}$.

\section{Experimental Results}
\label{section-experimental}

In this section, we briefly illustrate the implementation of the translation defined in Section \ref{section-ltl}.The test has been carried out to validate the correctness of rules in 
Table 2 from a practical point of view and it is, at the current state, only a proof-of-concept. We considered two workers, realizing the workflow of Figure~\ref{fig:first} and \ref{fig:second}, to prove the reachability of final states when a reconfiguration occur.
In particular, we impose that the reconfiguration process is activated 3 positions of time after the origin and that the first worker starts the execution at position 2 while the second one at position 4.

Checking the satisfiability of the translation of the workflow allowed us to verify whether there exists one execution with reconfiguration that satisfies the desired behaviour.
In other words, this means that the process of reconfiguration is correctly performed by the model.
To check the satisfiability of our LTL model and the property of termination, we exploit the Bounded Satisfiability Checking (BSC) \cite{PMS12} approach.
The key idea behind the BSC is to build a finite representation, of length $k$, of an infinite ultimately periodic LTL model of the form $\alpha\beta^\omega$, where $\alpha$ and $\beta$ are finite words over the alphabet $2^{AP}$.
Although the standard technique to prove the satisfiability of LTL formulae is based on the construction of B\"uchi automata \cite{vw}, the evidence has turned out that it is rather expensive in practice, even in the case of LTL (the size of the automaton is exponential with respect to the size of the formula) and that new techniques should be investigated.
BSC tackles the complexity of checking the satisfiability for LTL formulae by avoiding the unfeasible construction of the whole automaton.
By unrolling the semantics of the formula for a finite number $k>0$ of steps, BSC tries to build a bounded representation $\alpha\beta$, of length $k$, of an infinite ultimately periodic model for the formula of the form $\alpha\beta^\omega$.
\cite{BFRS11} proves that BSC problem for LTL and its extension is complete and that it can be reduced to a decidable Satisfiability Modulo Theory (SMT) problem.
It turns out that, in many cases, as our tests confirm, that the expensive construction of the B\"uchi automaton for the LTL formula can be avoided as it is possible to find finite models of small length in a very short time.

All our tests ware carried out by using the \zot{} \cite{PMP08} toolkit.
\zot{} is a Bounded Model/Satisfiability checker that takes as input specifications written in a variety of temporal logics, and determines whether they are satisfiable or not. Through this basic mechanism, it can perform verification of user-defined properties for the desired models, the workflow in our case. 

The methodology that we exploit to verify the translation and the reachability property for the workflow is the following.
Let $S$ be formula translating the workflow of Figure~\ref{fig:first} or  Figure~\ref{fig:second}.
If $S$ is fed to the \zot{} tool  "as is", \zot{} will look for one of its execution; if it does not find one (i.e., if the model is unsatisfiable), then the our translation is contradictory, hence it contains some flaws.
Now, let $P$ be a formula which formalizes the property of reachability we want to check on the workflow. 
If $S \land \neg P$ is unsatisfiable, this means that there is no execution that satisfies the workflow (i.e., $S$), that also satisfies $\neg P$, that is, that violates the property $P$. 
If no execution violates property $P$, then the latter actually holds for the workflow. 
If, on the other hand, $S \land \neg P$ is satisfiable, this means that there is at least one execution that satisfies both $S$ and $\neg P$; that is, there is at least one execution of the workflow that violates the property, so the property does not hold. 
If \zot{} determines that a formula is satisfiable, then the tool produces an execution that satisfies it that one can use to check the correcteness of the modelling.

\zot{} performs the checks outlined previously by encoding temporal logic formulae into the input language of various solvers. In particular, \zot{} supports two kinds of solvers: SAT solvers and SMT solvers. SAT solvers are capable of taking, as input, formulae written in propositional logic and determine whether they are satisfiable or not. SMT solvers do the same, but they accept, as input, formulae written in logics (fragments of First-Order Logic) that are richer than the simple propositional logic. Over the last few years, both SAT solvers and SMT solvers have made great strides in terms of their performances, so that they have become viable engines for fully automated logic-based verification approaches such as the one realized by \zot{}. In addition, most SAT/SMT solvers accept inputs written in a standard format, which makes them easily interchangeable. This is very useful, since different solvers implement different heuristics, and the "best" solver does not exist in absolute terms, but only on a model-by-model case.


\zot{} plugins implement the primitives that are used in \zot{} scripts to define the models to be analysed. Since the plugins are implemented as Lisp modules, the primitives are essentially Lisp declarations. Then, \zot{} scripts, which contain both the model to be analysed and the necessary commands to invoke the desired solver, are a collection of Lisp statements. 

Table \ref{tb:tests} shows the time required by the tool to check the reachability properties ($\F{s}$, where $s$ is activity \textit{Confirmation} or \textit{Archiving}), the memory occupation and the result, i.e. whether the property is satisfied or not.
All tests have been carried out on a 3.3 Ghz quad core PC with 16 Gbytes of Ram. 

The first formula refers to the reachability of the Accept state of the workflow in Figure \ref{fig:first}, represented by the first worker, which starts before the instant of reconfiguration. To show the impact of the model checker, we have changed the bound $k$, which is a user-defined parameter that corresponds to the maximal length of runs analysed by \zot{}. It corresponds to the number of steps needed to build the bounded representation of the model. The values chosen are $k=20$ and $k=100$.

The third formula refers to the reachability of the Accept state of the workflow in Figure \ref{fig:second}, represented by the second worker, which starts after the instant of reconfiguration. It is again been verified using two different values for the bound $k$. 

\setlength{\tabcolsep}{4pt}
\begin{table}[!hbt]
\begin{center}
\caption{Test results.}
\label{tb:tests}
\begin{tabular}{cccc}
Formula                            & Time (sec) & Memory (Mb) & Bound \\\hline
Accept state, Configuration 1 &  24.417        &   164    &  20\\
Accept state, Configuration 1   & 82.734      &   309    &  100 \\
Accept state, Configuration 2    & 28.873        &   187    &  20\\
Accept state, Configuration 2       &  94.74387        &   367    &  100
\end{tabular}
\end{center}
\end{table}

As it is expected, the time and memory consumption increases with the length of the bound of the analysed traces. Although the model is composed of a number of atomic propositions which is equal to the sum of the number of states and transitions of the two workflows (about 50), the needed time to perform the analysis is very low: this is due to the type of property analysed. For reachability properties the analysis is very faster, since the tool stops as soon as it finds a run that satisfies the formula; for example, deadlock detection analysis could take a long time, minutes or hours, as the tool must exhaustively analyse all possible runs. 

Furthermore, our tests have formally demonstrated the termination of workflows, before and after reconfiguration. 

We can conclude that it is feasible, using modern model checking tools such as \zot{}, to perform the formal verification of some simple properties, like reachability properties, in a limited amount of time. We left as future work the implementation and testing of more interesting properties, such as real-time properties. 

\section{Conclusions and future work}
\label{section-conclusions}

The objective of this paper is demonstrating how temporal logics and model checking are 
effective in proving properties of dynamic, reconfigurable and adaptable systems, in this
specific case business workflows. We proved workflow reconfiguration termination by providing 
a compilation of generic workflows into LTL and using \zot{} as a model checker. The idea of this
research stem by the observation that reconfigurability has not been thoroughly investigated in the
context of verification and the current state of the art fundamentally lacks logics and model 
checking paradigms able to describe and analyze complex modern systems. Therefore, with this 
research strand we are aiming towards a full exploitation of temporal logics and model checking
techniques for verification of self-adaptability, dynamically changing requirements and emergent 
behaviors. 

This simple case study is just a "proof of concept" to demonstrate the feasibility our
ideas, which now need to be challenged investigating a real industrial case study. In particular,
the workflow patterns here analyzed are limited with respect to a real scenario and therefore the nature of the encoded workflow and the details of the encoding need to be fully worked out. Workflow patterns as presented in \cite{wmp2003workflow} need to be investigated and eventually encoded. Once workflows are intended as graphs and transitions are treated like in our encoding, similarities emerge with the Petri Nets approach. In particular, we need to compare our work with Workflow Nets  \cite{Aalst:1997} and exploit possible synergies.

Future work aims at extending the current translation of workflows by using more expressive logics.
In particular, we would extend the basic definition of workflow to \textit{timed} workflow by adding timing constraints on activities and transitions. This extension allows us to constraint the duration of activities and of the execution flow with respect to time bounds that may vary with respect to certain environment parameters that are measured along the execution of the workflow. To model timed workflow we will exploit CLTLoc \cite{BRS13b}, that is an LTL based logic where atomic formulae are both atomic propositions and constraints over dense clocks.

Another field of future investigation, which is in the same direction of timed workflow, is the modelization of workflows where some activities take a neglected duration time respect to other ones. Infact, there are different real-time systems where the various activities of the workflow modeling them have a huge different time duration; for example, an activity which consists of the automatic reception of orders by e-mail has -- taken singularly --  a very short duration respect to the activity of checking the content of those orders, provided by humans in hours or days. Due to the different order of magnitudes in the duration of the two different activities, usually the first activity, with a neglected duration, is modeled as having a logical zero time duration. The zero-time modelization deals to a lot of different problematics, such as the rising of Zeno behaviours -- a sequence of infinite actions or activities executed in a finite or zero duration time, which is not a feasible behaviour of a real system -- or the fact that the "state" of the system -- which is represented by the set of atomic propositions which is true in a certain instant of time -- is not anymore a function of time, which is in a certain sense counterintuitive and may lead to logical contradiction. Some temporal logics adopt a time structure where more than one system state may be associated with a single time instant -- an example, reported in \cite{O89}, is the super-dense time model -- and therefore must provide distinct notations to refer to time and state change. In these notations, the progress of time and state evolution are fully decoupled, which is rather unnatural in most practical cases. To overcome this limitations, the work in \cite{FMMR12} introduce a new metric temporal logic called \textit{X-TRIO}, which exploits the concepts of \textit{Non-Standard Analysis} \cite{R96}, substituting each zero-time activity with an "infinitesimal" duration time activity, enriching the underline time model. As future work, we will study the way to "glue" together the CLTLoc logic with X-TRIO, adding to the first temporal logic the capability to model neglected duration time activities provided by the latter.

\centerline{} 

{\bf Acknowledgements.} I would like to acknowledge the help given by numerous colleagues, in particular: Anirban Bhattacharyya, Marcello Maria Bersani and Luca Ferrucci. This work has been supported by the Russian Ministry of education and science with the project "Development of new generation of cloudy technologies of storage and data control with the integrated security system and the guaranteed level of access and fault tolerance" (agreement: 14.612.21.0001, ID: RFMEFI61214X0001).
\bibliographystyle{plain}

\bibliography{bibliography}


\end{document}